\documentclass[12pt]{iopart}

\usepackage{graphicx}
\begin{document}

\newcommand{\cugeo}{CuGeO$_3$\,}
\newcommand{\dcugeo}{Cu$_{1-x}$Cd$_x$GeO$_3$\,}

\title[Neutron and X-ray Scattering Studies of \dcugeo] {Neutron and X-ray Scattering Studies of the Lightly-Doped
Spin-Peierls System \dcugeo}

\author{S. Haravifard$^{1}$, K.C. Rule$^{1}$, H.A. Dabkowska$^{1}$ and B.D. Gaulin$^{1,2}$}
\address{$^{1}$Department of Physics and Astronomy, McMaster University,
Hamilton, Ontario, L8S 4M1, Canada}
\address{$^{2}$Canadian Institute for Advanced Research, 180 Dundas St. W.,
Toronto, Ontario, M5G 1Z8, Canada}
\author{Z. Yamani$^{3}$ and W.J.L. Buyers$^{2,3}$}
\address{$^{3}$Canadian Neutron Beam Centre, NRC, Chalk River Laboratories, Chalk River, Ontario, K0J 1J0, Canada}
\date{\today}
\ead{gaulin@mcmaster.ca}

\begin{abstract}
Single crystals of the lightly-doped spin-Peierls system \dcugeo
have been studied using bulk susceptibility, x-ray diffraction,
and inelastic neutron scattering techniques.  We investigate the
triplet gap in the magnetic excitation spectrum of this quasi-one
dimensional quantum antiferromagnet, and its relation to the
spin-Peierls dimerisation order parameter.  We employ two
different theoretical forms to model the inelastic neutron
scattering cross section and $\chi''$({\bf Q}, $\omega$), and
show the sensitivity of the gap energy to the choice of
$\chi''$({\bf Q}, $\hbar\omega$). We find that a finite gap exists at the
spin-Peierls phase transition.
\end{abstract}

\submitto{J. Phys.:Condensed Matter}

\section{Introduction}
Low dimensional quantum magnets\cite{Review} which display
collective singlet ground states are very topical, due to the
exotic low temperature properties they display, as well as their
relation to high temperature superconductivity\cite{HighTC}.
Quasi-two dimensional S=1/2 systems such as the Shastry-Sutherland
system SrCu$_2$(BO$_3$)$_2$
exist\cite{ShastrySutherland,Miyahara,Kageyama1999,Gaulin},
wherein orthogonal Cu$^{2+}$ dimers are arranged on a square
lattice.  This material displays a collective singlet ground
state, relatively dispersionless triplet excitations and multiple
triplet bound excited states.  Quasi-one dimensional quantum
magnets are more common, with S=1/2 chains based on organic
molecules, such as TTF-CuBDT\cite{TTFa,TTFb} and
MEM-(TCNQ)$_2$\cite{MEM,Lumsden1999}, based on Cu$^{2+}$ (3d$^9$),
such as
CuGeO$_3$\cite{CGO_review,Hase1993a,Hirota,Pouget,Kamimura,Nishi,Fujita,Regnault1996,Lumsden1998a,Birgeneau},
and most recently based on Ti$^{2+}$(3d$^1$), such as
TiOCl\cite{Seidel, Imai, Clancy} and TiOBr\cite{Sasaki}.  These
materials undergo spin-Peierls phase transitions to a singlet
ground state as the temperature is lowered. Related phenomena
occurs in quasi-one dimensional quantum magnets with S=1 chains,
such as NENP and CsNiCl$_3$ \cite{NENP,Buyers}, which enter a
Haldane singlet phase at low temperatures.

CuGeO$_3$ was the first inorganic spin-Peierls system to be
discovered. The singlet ground state associated with CuGeO$_3$
below its spin-Peierls phase transition of T$_{SP}$$\sim$ 14.1 K
has been well
studied\cite{CGO_review,Hase1993a,Hirota,Pouget,Kamimura,Nishi,Fujita,Regnault1996,Lumsden1998a,Birgeneau}.
Such a system is characterized by uniform chains of S=1/2 moments
at high temperatures, which dimerise at low temperature to allow
singlets to form. This phase transition breaks translational
symmetry, and a singlet-triplet gap is introduced into its
magnetic excitation spectrum at its magnetic zone centre. It
possesses a much higher magnetic moment density than the
pre-existing organic spin-Peierls
systems\cite{TTFa,TTFb,MEM,Lumsden1999}, and it can be grown in
large single crystal form by several different growth techniques.
This has enabled detailed neutron scattering studies of the
spin-Peierls ground state and its excitations\cite{Regnault1996}.
CuGeO$_3$ can also be grown in the presence of impurities, and
studies of doped-CuGeO$_3$ have revealed the sensitivity of the
spin-Peierls ground state to different types of
impurities\cite{CGO_review, Lumsden1998b}. In particular they have
revealed a remarkably rich temperature-impurity concentration
phase diagram in which antiferromagnetic long range order coexists
with either a dimerised or uniform structure at sufficiently low
temperatures\cite{Hase1993b,Oseroff,Lussier1995,Sasago,Renard,Regnault1995,Poirier,Masuda}.
This occurs for both non-magnetic
Zn$^{2+}$\cite{Hase1993b,Oseroff,Lussier1995,Sasago}and
Mg$^{2+}$\cite{Masuda} substituting for Cu$^{2+}$, as well as for
Si$^{4+}$\cite{Oseroff,Renard,Regnault1995,Poirier} substituting
for Ge$^{4+}$.

Most of the work on impurities in CuGeO$_3$ has employed dopants
which possess a similar or a smaller ionic radius than that of the
host ion which they seek to replace.  Zn$^{2+}$, Mg$^{2+}$, and
Cu$^{2+}$ have ionic radii of 0.74 $\AA$, 0.66 $\AA$, and 0.72
$\AA$, respectively.  However, some work has also been done on low
concentration substitution\cite{Lumsden1998b} of Cu$^{2+}$ with
Cd$^{2+}$, whose ionic radius is much bigger, 0.97 $\AA$, than
that of Cu$^{2+}$. The difference in ionic radii severely limits
the solubility of Cd in CuGeO$_3$; nonetheless small single
crystals of \dcugeo with x $\le$ 0.002 were grown and
studied\cite{Lumsden1998b}.  This previous
study\cite{Lumsden1998b}, on small single crystals grown from a
flux, showed little change in T$_{SP}$, and no coexisting
antiferromagnetism at the low Cd concentrations and base
temperature that could be achieved.  However, interestingly, the
critical properties of the spin Peierls phase transition changed
from three dimensional universality to mean field behaviour on
doping with Cd.

One interesting dimension of the spin-Peierls problem is the
relation between the singlet-triplet gap in the spin excitation
spectrum, and the order parameter for dimerisation.  Cross and
Fisher\cite{CrossFisher} originally argued for the power law
relation $\Delta$(T) $\sim$ ($\delta$ (T))$^{\nu}$ with $\nu$=2/3.
The discovery of the spin-Peierls state in CuGeO$_3$ has allowed
this relationship to be tested directly using inelastic and
elastic neutron scattering to measure the temperature dependence
of the  gap energy, $\Delta$(T), and the square of the order
parameter for the spin-Peierls dimerisation $\delta^2$.  We report
here inelastic neutron scattering measurements of the temperature
dependence of the magnetic excitation spectrum at the magnetic
zone centre, and of the x-ray diffraction measurements of the
superlattice Bragg peak intensity, in a new single crystal of
lightly doped \dcugeo (x $\le$ 0.002). These results show that the
simple power law relation between the gap and the dimerisation
order parameter is not obeyed, and that a finite triplet gap
exists at the spin-Peierls phase transition itself.

\section{Experimental Details}

A single crystal of \dcugeo with x $\le$ 0.002 was grown by
the self-flux method in a floating zone image furnace. The crystal was
grown at a rate of $\sim$ 5-8 mm/hour with an oxygen pressure of
47 kPa. Earlier experience\cite{Lumsden1998b} on flux-grown
\dcugeo indicated a low solubility of Cd in the \cugeo host.  For
that reason, naturally occurring Cd was used in the crystal
growth, even though the crystals were intended for neutron
scattering studies, and Cd has a high neutron absorption cross
section. Initial neutron diffraction measurements on the sample
showed strong Bragg scattering, from a high quality crystal that
was single throughout its volume. Its approximate dimensions were
of 30 mm long by 5 mm in diameter, and mosaic spread was less than
0.4 degree. These measurements confirmed that the crystal was
orthorhombic with lattice parameters within error the same as the
pure material; \textit{a}= 4.81{\AA}, \textit{b}= 8.47{\AA} and
\textit{c}= 2.94{\AA} at 4K.

X-ray diffraction measurements were performed on a small single
crystal cut from the large crystal used in the neutron scattering
measurements.  The crystal was mounted on the cold finger of a
closed cycle refridgerator and aligned within a Huber four circle
goniometer. The measurements with a rotating anode Cu-K$\alpha$
x-ray source and a pyrolitic graphite monochromator were
performed at
temperatures from 6.5 K to 14.5 K with a temperature stability of
$\sim$ 0.005 K. The primary purpose of these measurements was to
precisely study the critical properties of the spin-Peierls order
parameter, as measured by the temperature dependence of the {\bf
Q}=($\frac{1}{2}$,5,-$\frac{1}{2}$) superlattice Bragg peak
intensity, and to determine the critical exponent, $\beta$.

\begin{figure}[h]
\centering \resizebox{4.6in}{3.7in}{\includegraphics{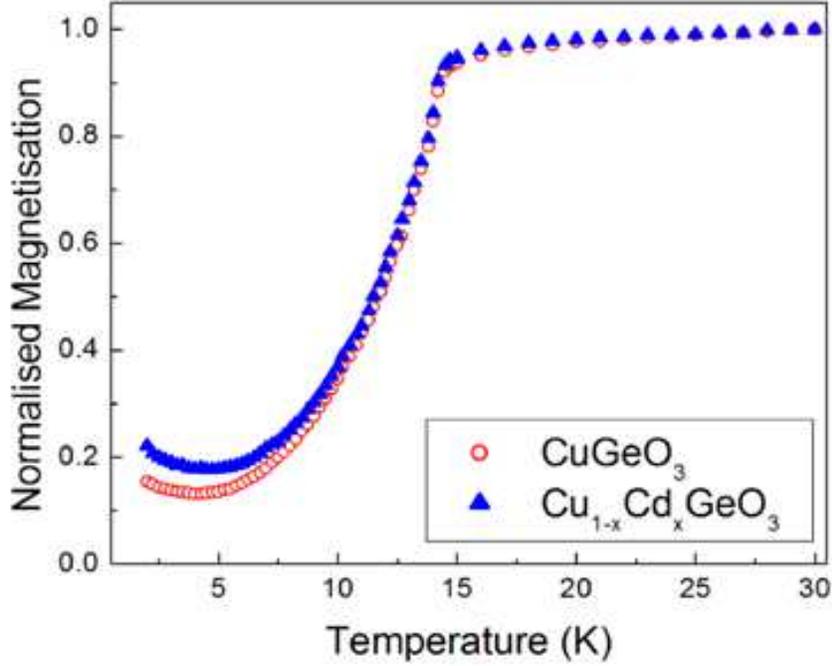}}
\caption{SQUID dc-susceptibility measurements (with 1000G applied
magnetic field) on CuGeO$_{3}$ and Cu$_{1-x}$Cd$_{x}$GeO$_{3}$
with x $\leq$ 0.002 are compared.} \label{Figure 1}
\end{figure}

Another small piece of crystal was cut off and used for magnetic
characterization with SQUID magnetometry.  The characteristic
falloff of the dc susceptibility signifying T$_{SP}$ near 14.1 K
was observed. Figure 1 shows the comparison of the normalised
susceptibility measurement for both \cugeo and \dcugeo samples as
a function of temperature. It is clear from this data that the
susceptibility of the doped sample is very similar to that of the
pure material.  At temperatures above T$_{SP}$, the susceptibility
of both samples show a broad maximum characteristic of short
range, quasi-one-dimensional correlations. Below 10 K the \dcugeo
susceptibility is $\sim$ 20$\%$ larger than that of the \cugeo
sample, indicating that Cd impurities are indeed present in the system.
They have the effect of freeing-up individual spins near the impurities,
thereby increasing the susceptibility.

Elastic and inelastic neutron scattering measurements were
performed on the large single crystal of \dcugeo at the Canadian
Neutron Beam Centre, Chalk River, using the N5 triple axis
spectrometer.  The crystal was mounted in a $^3$He cryostat with
its (0,K,L) plane coincident with the horizontal scattering plane,
such that wavevectors near the {\bf Q}=(0,1,$\frac{1}{2}$)
magnetic zone centre could be accessed.  The measurements were
made with pyrolytic graphite as both monochromator and analyser
crystals, a fixed final neutron energy of 14.7 meV, and with two
pyrolytic graphite filters in the scattered beam to reduce higher
order contamination. Soller slits determined the horizontal
collimation and the resulting horizontal and vertical divergences
of the beam were [38,36,36,212] and [58,73,146,636] respectively,
in minutes of arc, using the convention [source-monochromator,
monochromator-sample, sample-analyser, analyser-detector].

Elastic neutron scattering measurements were performed at the
magnetic zone centre, {\bf Q}=(0,1,$\frac{1}{2}$), to search for
impurity-induced antiferromagnetic ordering at T=0.32 K. No
evidence for magnetic ordering was found.

The lack of change in T$_{SP}$ in \dcugeo as compared with \cugeo,
as well as the absence of magnetic order at T=0.32 K, can be used
to set an upper limit for the Cd concentration in the single
crystal sample of \dcugeo. Assuming that the
Cu$_{1-x}$Mg$_x$GeO$_3$ phase diagram\cite{Nishi2000} is
applicable to \dcugeo, at least at low doping concentrations, an
upper limit of x $\leq$ 0.002 can be set.

\section{Experimental Results and Analysis}

\subsection{X-ray Diffraction}

We measured the temperature dependence of the $\bf
Q$=($\frac{1}{2}$,5,-$\frac{1}{2}$) superlattice Bragg peak
intensity, as shown in figure 2 for temperatures close to
T$_{SP}$. This peak arises from the dimerisation pattern within
the spin-Peierls state in \cugeo, and its amplitude is
proportional to the square of the order parameter.

As was done previously to examine the critical properties of doped
\cugeo\cite{Lumsden1998b}, this peak intensity as a function of
temperature was fit to a modified power law as shown in equation
1. This modified power law includes a correction to scaling
term\cite{Aharony}, with the correction to scaling exponent
${\eta}$ set to its expected value of 0.5 and $t=\frac{T_{SP}-T}{T_{SP}}$.

\begin{equation}
    I = I_{0}t^{2\beta}\left(1+At^{\eta}\right)+Background
\end{equation}

\begin{figure}[h]
\centering \resizebox{4.8in}{3.7in}{\includegraphics{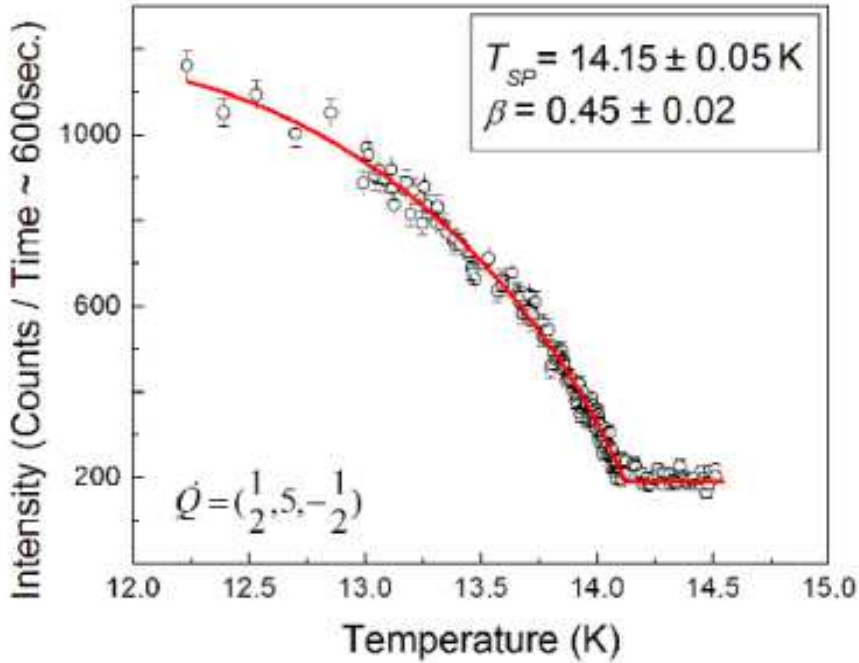}}
\caption{X-ray scattering measurements of the superlattice Bragg
intensity at $\bf Q$=($\frac{1}{2}$,5,-$\frac{1}{2}$) are shown as
a function of temperature.  The solid line shows a fit of this
temperature dependence to critical behaviour described in equation
1. We observe mean field-like behaviour, consistent with earlier
measurements on small flux grown \dcugeo single crystals
\cite{Lumsden1998b}.} \label{Figure 2}
\end{figure}

The solid line in figure 2 shows the fit of equation 1 to the data
and clearly this expression describes the data very well for
temperatures close to T$_{SP}$.  The fit gives T$_{SP}$ = 14.15
$\pm{0.05}$ K and a critical exponent $\beta = {0.45} \pm {0.02}$.
This value is close to the mean-field value of $\beta = {0.5}$,
and is much larger than the values for $\beta(\sim0.33)$ from
three dimensional universality  \cite{Collins,Newman} that are
known to characterize both pure \cugeo\cite{Lumsden1998a,
Birgeneau} and lightly doped \cugeo in which the dopants possess
similar ionic radii to the host ions they
replace\cite{Lumsden1998b}.

This mean field result is similar to that found by Lumsden et al.\cite{Lumsden1998b}
in which the critical behaviour of lightly doped single crystals of \dcugeo
grown by the flux method also showed mean field critical exponent $\beta$ values.
These results establish that some Cd impurities are present in the crystal, and also
provide a quantitative form for the spin-Peierls order parameter as a function of temperature,
which can then be compared to the temperature dependence of the triplet gap in the
excitation spectrum obtained from inelastic neutron scattering.

\subsection{Neutron Scattering}

Constant-\textbf{Q} inelastic neutron scattering scans were
performed at the magnetic ordering wavevector {\bf Q$_{0}$}=(0,1,$\frac{1}{2}$) in order to observe the temperature
dependence of the triplet excitations at the magnetic zone centre.
The energies of the triplet excitations disperse with wavevector
due to the three dimensional nature of the magnetic system.  Near
the ordering wavevector, this dispersion\cite{Regnault1996} varies
with the relative wave vector
\textbf{q}=\textbf{Q}-\textbf{Q}$_{0}$, as:

\begin{equation}
  \Delta_{\textbf{q}} = \sqrt{\Delta^{2} + (v_{a}q_{a})^{2} + (v_{b}q_{b})^{2} + (v_{c}q_{c})^{2}}
\end{equation}
where $\Delta$ is the minimum triplet excitation energy
or gap energy. $q_{a}$, $q_{b}$ and $q_{c}$ are reduced wave
vectors expressed in r.l.u. where:\\

$v_{a}q_{a} = (\Delta E)_{a}\sin(\pi q_{a})$  and  $(\Delta E)_{a}
\approx 1.66 meV$

$v_{b}q_{b} = (\Delta E)_{b}\sin(\pi q_{b}/2)$  and  $(\Delta
E)_{b} \approx 5.3 meV$

$v_{c}q_{c} = c_{0} q_{c}$ and  $c_{0} = 80 meV$\\

\begin{figure}[h]
\centering \resizebox{5.0in}{4.0in}{\includegraphics{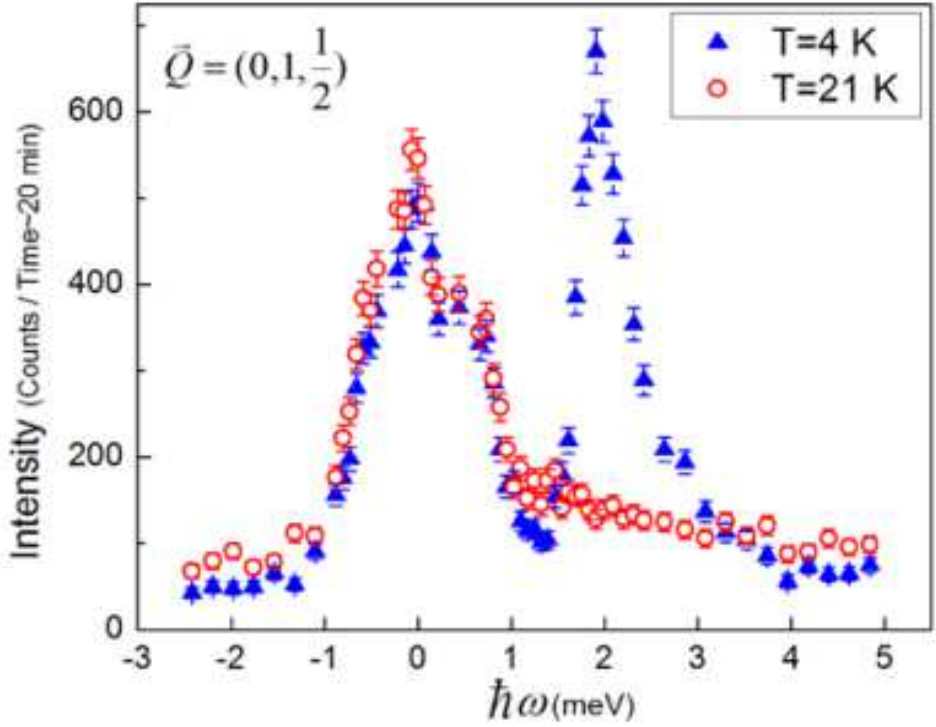}}
\caption{Constant-\textbf{Q} inelastic neutron scattering scans at
the magnetic zone centre $\bf Q$=(0,1,$\frac{1}{2}$), taken well
below T$_{SP}$ at T=4 K and well above T$_{SP}$ at T=21 K.}
\label{Figure 3}
\end{figure}

Representative data at 4 K and 21 K, well below and well above
T$_{SP}$ respectively, are shown in figure 3, and the low
temperature singlet-triplet gap of $\Delta$ $\sim$ 2 meV is
identified in the 4 K data.  At 21 K, the triplet excitation is
completely absent and the finite energy peak in the inelastic
scattering has been replaced with a weak continuum of scattering
from quasi-elastic energies, out to the end of the scan, 4.8 meV.
We note that the triplet excitation at low temperatures
exhibits an asymmetric tail to the high energy side.

Figure 4 shows the temperature dependence of the quasi-elastic
scattering at $\hbar\omega$=0.41 meV and at $\bf
Q$=(0,1,$\frac{1}{2}$).  It shows a ``critical" regime which
extends from $\sim$ 12 K to $\sim$ 21 K within which quasi-elastic
scattering is significantly-enhanced compared with either
lower or higher temperatures.  We wish to isolate the triplet
excitation at $\bf Q$=(0,1,$\frac{1}{2}$), from the incoherent
elastic scattering as well as from the background scattering, and
thereby determine the gap energy, $\Delta$, as a function of
temperature. This requires a background subtraction for which we
have two options. We can use the low temperature scattering at T=4
K - suitably modified to exclude the resolution - limited triplet
excitation, or we can use the scattering at 21 K.  Each of these
has advantages. For the T=4 K data the triplet excitation is sharp
in energy, and so can be cleanly separated from the remaining
scattering, comprised of incoherent elastic scattering from the
sample, and energy-independent background scattering from fast
neutrons. However the use of the T=4 K data set as a background
does not recognise that inelastic scattering persists above
T$_{SP}$, albeit in the form of a weak, quasi-elastic spin
excitation spectrum for which the T=21 K data set is
characteristic.  In what follows, we employ both a suitably
modified T=4 K data set (LT background) as well as the T=21 K data
set (HT background) as the background data set to be subtracted
from the signal so as to accurately estimate the scattering from
the triplet excitation alone. This will allow us to examine the
sensitivity of the gap, $\Delta (T)$, to the method of background
scattering estimation.

\begin{figure}[h]
\centering \resizebox{4.3in}{3.5in}{\includegraphics{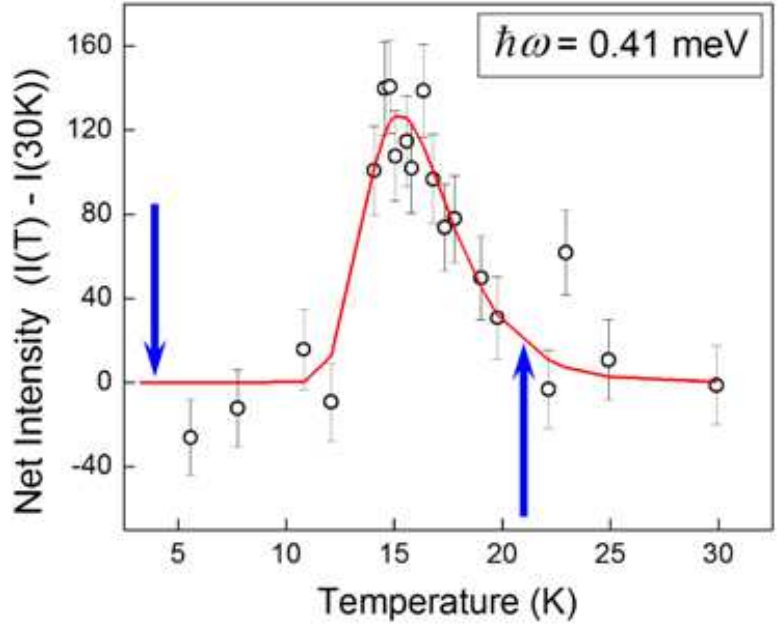}}
\caption{Net intensity of the neutron scattering observed at the
magnetic zone centre, $\bf Q$=(0,1,$\frac{1}{2}$) and an energy
transfer of 0.42 meV is shown.  The solid line is a guide to the
eye and the vertical arrows indicate the temperatures at which the
low temperature and high temperature background data sets were
taken.} \label{Figure 4}
\end{figure}

Figure 5 shows representative constant-{\bf Q} scans at
(0,1,$\frac{1}{2}$), for which a high or low temperature data sets
have been subtracted from the scans at temperatures ranging from
$\sim$ 0.7 T$_{SP}$ (10 K) to $\sim$ 1.1 T$_{SP}$ (15 K).  This
data was fit to two different forms for S({\bf Q}, $\omega$) with
the intention of determining the temperature dependence of the gap
energy, $\Delta$, and the inverse lifetime, $\Gamma$ of the
triplet excitations.  However a qualitative examination of the
data in figure 5 shows a substantial, well defined inelastic peak
to exist at $\sim$ 1.6 meV or $\sim$ 1.8 meV and T=13.83 K $\sim$
0.98 T$_{SP}$, depending on whether the low temperature (LT) or
high temperature (HT) data set is used as a background.  One would
qualitatively conclude therefore that the gap remains finite at
T$_{SP}$ in this sample of \dcugeo, a result that is borne out by
a quantitative analysis of the excitation spectrum, discussed
below.

\begin{figure}[h]
\centering \resizebox{4.7in}{7.6in}{\includegraphics{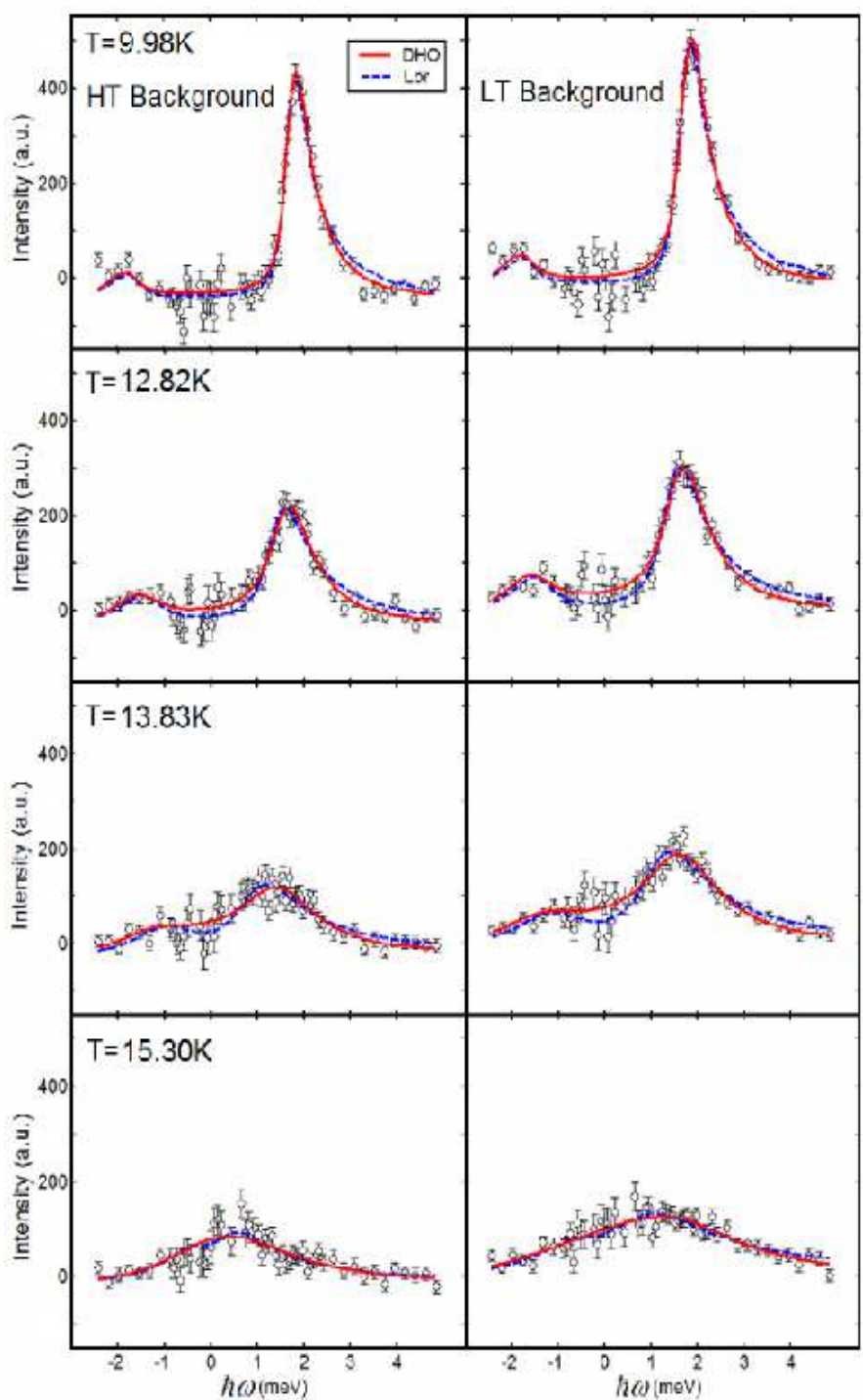}}
\caption{Representative inelastic spectrum, below and above
T$_C$=14.15 K, and at the magnetic zone center {\bf Q}=
(0,1,$\frac{1}{2}$) are shown, using the high temperature data set
as background (left hand panels), and the low temperature data set
as background (right hand panel).  The lines through the data show
fits of the spectra to a Lorentzian and a DHO form of S({\bf Q},
$\omega$), as described by equations 3 and 4, respectively. The
DHO model is clearly superior especially at large energies.}
\label{Figure 5}
\end{figure}

The inelastic spectra, shown in figure 5, were fit to two models
of S({\bf Q}, $\omega$), each of which was convolved with the
four-dimensional instrumental resolution function.  The finite
resolution of the measurement combines with the dispersion of the
triplet excitations to higher energies at wavevectors away from
the magnetic zone centre, equation 2, and results in the asymmetry
of the triplet lineshape, with a high energy tail.  This is
accounted for within our resolution convolution, where we employed
the spin wave velocities (see equation 2) determined
previously\cite{Regnault1996}.

The first Lorentzian  model was employed by Regnault et
al\cite{Regnault1996}, in their analysis of the temperature
dependence of the triplet excitation energy near the magnetic zone
centre in pure \cugeo. The Lorentzian (Lor) profile is given by:

\begin{equation}
   S_{L}(\mathbf{Q},\omega)\sim\frac{\omega}{1-exp({-\omega}/{kT})}
  \left[\frac{\Gamma_{L}}{(\omega-\Delta_{L})^{2}+\Gamma_{L}^{2}}+\frac{\Gamma_{L}}{(\omega+\Delta_{L})^{2}+\Gamma_{L}^{2}}\right]
\end{equation}

The second  model was a damped harmonic
oscillator (DHO) given by:

\begin{equation}
    S_{D}(\mathbf{Q},\omega)\mathrel=\frac{\chi_{0}\Delta^{2}\pi^{-1}}{1-exp({-\omega}/{kT})}
    \left[\frac{2\omega\Gamma}{(\omega^{2}-\Delta^{2})^{2}+ 4\omega^{2}\Gamma^{2}}\right]
\end{equation}

where $\chi_{0}$ is the static
susceptibility at \textbf{Q}. For small damping, the relation
between the gap energy in the DHO and Lor models is:

\begin{equation}
  \Delta_L^{2} = \Delta^{2}- \Gamma^{2}
\end{equation}

The results of fitting the data to the Lor model and the DHO model
are shown as the solid and dashed lines in figure 5. Both models
are reasonable descriptors of the data. However, the DHO model is
a better descriptor as its goodness-of-fit parameter, $\chi^2$, is
typically 10 to 40$\%$ lower than for the Lor model at all
temperatures. This is because the Lor spectrum, used earlier
\cite{Regnault1996}, falls off too slowly with $\omega$; indeed its
integral in frequency is divergent. We conclude
that the inelastic scattering is best described using the DHO
form for S({\bf Q}, $\omega$), equation 4.

The values of the gap energy, $\Delta$, and inverse lifetime,
$\Gamma$, of the triplet excitations extracted from this analysis
are plotted as a function of temperature in figure 6.  The top
panel shows the parameters resulting from an analysis of the data
using the HT background, while the bottom panel shows the
parameters relevant to the LT background.  As can be seen from
figure 6, while the background data set used influences the
details of the fit parameters, it does not affect the overall
trends and general features of the temperature dependence of the
gap and inverse lifetime of the triplet excitations.

\begin{figure}[h]
\centering \resizebox{3.8in}{6.2in}{\includegraphics{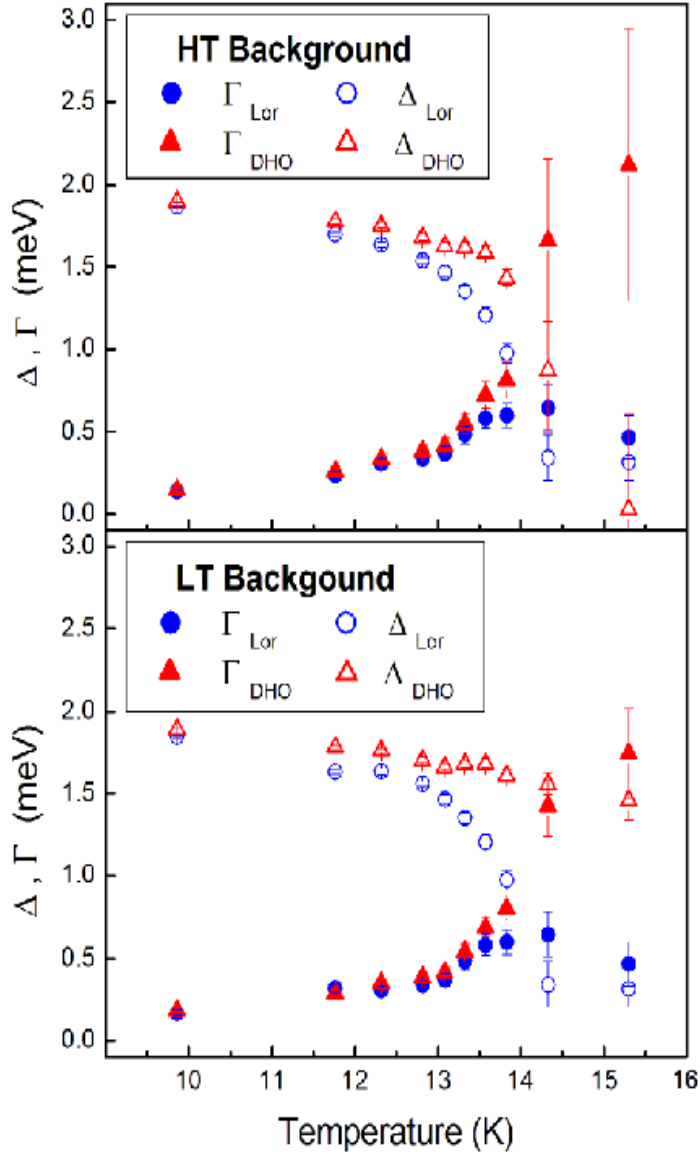}}
\caption{The temperature dependence of the energy gap, {$\Delta$},
and inverse lifetime, {$\Gamma$}, of the triplet excitation is
shown.  The upper panel shows the fit parameters for both the
Lorentzian and DHO models with the high temperature background
subtraction, while the lower panel shows the the parameters
extracted using the low temperature background subtraction.}
\label{Figure 6}
\end{figure}

\section{Discussion}

Our analysis, employing two different forms of S({\bf Q}, $\omega$)
and two different background subtractions, results in four forms of
the gap energy, $\Delta$, and inverse lifetime, $\Gamma$, as a
function of temperature, which can then be compared with
theoretical expectations.  These are plotted as a function of
temperature in figure 6, where the top panel shows the parameters
arising from use of the HT background, and the bottom panel shows
those arising from use of the LT background.  As can be seen the
gap energy, $\Delta \sim$ 2 meV at 4 K, is independent of both the
form of S({\bf Q}, $\omega$) and the details of the background,
provided the lifetime of the triplets is sufficiently long, as it
is below T$\sim$ 10 K. Above $\sim$ 10 K, differences between the
fitted gap energies progressively increase as the energy width
of the excitations, and hence the inverse lifetimes, become
larger. However in all four gap vs temperature plots shown in
figure  6, the gap energy, $\Delta$, does not appear to go to zero
at T$_{SP}\sim$ 14.15 K. Rather the phase transition occurs where
the gap energy, $\Delta$, and the energy width or inverse lifetime
of the excitation ,$\Gamma$, cross.

Figure 7 shows the gap energy, $\Delta$, plotted as a function of the
spin-Peierls order parameter as determined from the x-ray
scattering determination of the temperature dependence of the
superlattice Bragg peak intensity shown in figure 2.  This net
intensity is proportional to the square of the order parameter,
and consequently we have plotted the square root of the net
intensity on the x-axis of figure 7.  For reference, an x-axis
label has been added to the top of figure 7 to denote the actual
temperature.  The top panel of figure 7 shows the analysis using
the DHO form of S({\bf Q}, $\omega$), while the bottom panel shows
that using the Lorentzian form.

\begin{figure}[h]
\centering \resizebox{4.2in}{6.2in}{\includegraphics{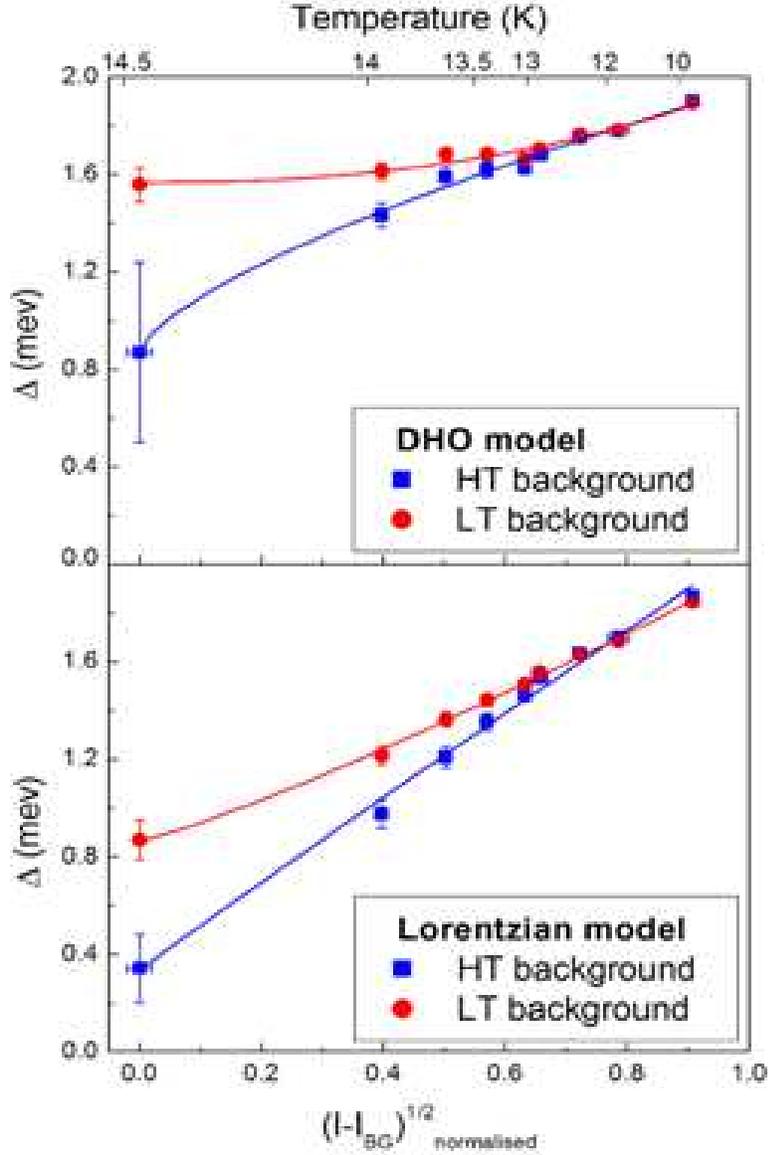}}
\caption{The gap energy, {$\Delta$}, as a function of temperature
is correlated with the corresponding spin-Peierls dimerisation
order parameter, taken as the square-root of the net x-ray
scattering intensity at the superlattice reflection ${\bf
Q}$=($\frac{1}{2}$,5,-$\frac{1}{2}$).  The upper panel shows the
parameters resulting from the DHO form of S({\bf Q}, $\omega$),
while the lower panel shows the parameters obtained from the
Lorentzian form of S({\bf Q}, $\omega$). The lines in the figure
are the results of fitting the data to $\Delta
(T)$$=$$\Delta_0$+$\delta (T)$$^\nu$, with $\Delta_0$ a free
parameter and also with $\Delta_0$ set equal to zero, as described
in the text.} \label{Figure 7}
\end{figure}

The systematic dependence of the gap on the form of S({\bf Q},
$\omega$) and the details of the background can be seen in figures
6 and 7.  As T$_{SP}$ is approached, the DHO form of S({\bf Q},
$\omega$) produces a higher value of the gap energy as compared
with the Lorentzian form.  For either form of S({\bf Q},
$\omega$), the use of the LT background results in a higher gap
energy near $T_{SP}$, as compared to when the HT background is
used.

As previously discussed, theoretical expectations exist for the
relation between the gap energy and the spin-Peierls order
parameter; $\Delta (T)$$\sim$$\delta (T)$$^\nu$ with
$\nu$=$\frac{2}{3}$.  This argument was originally made by Cross
and Fisher\cite{CrossFisher} in the context of the spin-Peierls
transition in TTF-CuBDT\cite{TTFa,TTFb}.  We have therefore fit
the data shown in figure 7 to $\Delta (T)$$=$$\Delta_0$+$\delta
(T)$$^\nu$, with $\Delta_0$ a free parameter and also with
$\Delta_0$ set equal to zero.  This latter case, with the gap
going to zero at T$_{SP}$, is consistent with the original
theoretical expectation\cite{CrossFisher}. The results of fitting
our data to this expression are given in table 1, for all four
data sets (DHO and Lor forms of S({\bf Q}, $\omega$), and both HT
and LT backgrounds).  As can be seen, the fits give a finite value
of $\Delta_0$, the gap energy at T$_{SP}$, unless it is
constrained to be zero.  Only one of the four combinations of
S({\bf Q}, $\omega$) and background (Lor and HT background) give
behaviour which is roughly consistent with $\Delta
(T)$$\sim$$\delta (T)$$^\nu$ with $\nu$=$\frac{2}{3}$, and this
case gives a finite gap at T$_{SP}$ of 0.34 $\pm$ 0.02 meV. The
fits using the DHO form of S({\bf Q}, $\omega$), which allows for
the better quality description of the inelastic neutron scattering
spectra, give either very low values of the exponent $\nu$, or
non-zero values of the gap at T$_{SP}$ ranging from $\sim$ 0.4 to
0.75 $\times$ the zero temperature value of the gap. We therefore
conclude that the predicted relation $\Delta (T)$$\sim$$\delta
(T)$$^\nu$ with $\nu$=$\frac{2}{3}$ is not obeyed in \dcugeo, and
that the gap is finite at T$_{SP}$.

A question arises as to what role doping plays in this behaviour?
Systematic studies of the Cu$_{1-x}$Mg$_x$GeO$_3$ have shown a
``pseudogap" temperature regime to exist above T$_{SP}$ for the
low dopant concentrations which allow a spin-Peierls transition to
occur\cite{Nishi2000}. This temperature regime is bordered from
below by the appearance of long range spin-Peierls order, and from
above by signatures indicative of the presence of a gap, such as a
suppression in the susceptibility.  This pseudogap regime broadens
in temperature with increasing doping until the spin-Peierls state
is lost altogether beyond x$\sim$ 0.03 in Cu$_{1-x}$Mg$_x$GeO$_3$.
The low doping level present in the \dcugeo sample studied here,
and the observation of pseudogap-like behaviour in
Cu$_{1-x}$Mg$_x$GeO$_3$, suggest that the finite gap at T$_{SP}$
may be intrinsic to pure CuGeO$_3$ as well. Surprisingly, in a
previous study\cite{CrossFisher} the temperature dependence of
$\Delta (T)$ for pure CuGeO$_3$ was found to be consistent with
$\Delta (T)$$\sim$$\delta (T)$$^\nu$ with $\nu$=$\frac{2}{3}$.
However these earlier measurements focused on the triplet
excitations at a wavevector slightly displaced from the
dimerization zone centre, and employed the less satisfactory
Lorentzian form for S({\bf Q}, $\omega$) only. It may be of
interest to revisit this problem in pure CuGeO$_3$.

It is notable that pseudo-gap behaviour has also been observed in
the unconventional spin-Peierls material TiOCl\cite{Imai,Clancy}.
This material exhibits both a low temperature dimerization into a
singlet ground state below T$_{SP1}$ and an intermediate
temperature phase characterized by an incommensurate structural
distortion.  Above this phase transition, a uniform phase exists
which displays characteristics of a finite gap and the NMR
signature for this pseudogap is maintained to $\sim$ 1.3
T$_{SP2}$\cite{Imai}.

\begin{table}
\caption{\label{math-tab2}Values of the exponent {$\nu$} used to describe the power law relationship between {$\Delta$} and the order parameter}
\begin{tabular*}{\textwidth}{@{}l*{15}{@{\extracolsep{0pt plus12pt}}l}}
\br
 &HT background&&LT background&\\
 &{$\nu$}&  {$\Delta_{0}$}&{$\nu$}& {$\Delta_{0}$}\\
\mr
DHO&0.43 $\pm$ 0.06 &0.85 $\pm$ 0.04&1.0 $\pm$ 0.4 &1.54 $\pm$ 0.04\\
Lorentzian&0.67 $\pm$ 0.02&0.34 $\pm$ 0.02&0.82 $\pm$ 0.05 &0.88 $\pm$ 0.02\\
\br
\br
DHO&0.30 $\pm$ 0.03 &0 &0.13 $\pm$ 0.03 &0\\
Lorentzian&0.81 $\pm$ 0.04&0&0.48 $\pm$ 0.01 &0\\
\br
\end{tabular*}
\end{table}

\section{Conclusions}
Inelastic neutron scattering and x-ray diffraction measurements
were carried out on a lightly doped sample of \dcugeo.  These
x-ray diffraction measurements of the superlattice Bragg intensity
below T$_{SP}$ confirmed the mean field behavior of the spin-Peierls
phase transition in this new large single crystal of \dcugeo grown by
floating zone image furnace techniques.  This result is consistent
with earlier critical scattering measurements on small single
crystals of \dcugeo grown by flux techniques\cite{Lumsden1998b}.

The order parameter measurements as a function of temperature were
correlated with inelastic neutron scattering measurements of the
singlet-triplet energy gap in this singlet ground state system,
for the purpose of testing the relationship between the gap energy
and the spin-Peierls order parameter.  We investigated the
sensitivity of the gap energy extracted from an analysis of the
inelastic scattering, to the form used to model S({\bf Q},
$\omega$) and to the form of the background scattering.  This
analysis showed the inelastic scattering to be best described
using a DHO form for S({\bf Q}, $\omega$). We find that the energy
gap remains finite at T$_{SP}$, as opposed to going to zero, as
might have been anticipated on the basis of earlier theoretical
expectations\cite{CrossFisher}.

We hope that this result on lightly doped \dcugeo can inform and motivate further work on spin-Peierls
and other singlet ground state systems, and shed light on the formation of the triplet gap at and above T$_{SP}$.

\ack We acknowledge the expert technical assistance provided by
the staff at NRC-CNBC, Chalk River. This work was supported by
NSERC of Canada.
\\

\end{document}